# Interaction between nearly hard colloidal spheres at an oil-water interface

Iain Muntz,[1] Franceska Waggett,[2] Michael Hunter,[3,1] Andrew B. Schofield,[1] Paul Bartlett,[2]
Davide Marenduzzo,[1] and Job H. J. Thijssen [1,*]

[1]*SUPA School of Physics and Astronomy, The University of Edinburgh, Edinburgh, EH9 3FD, Scotland, United Kingdom*
[2]*School of Chemistry, University of Bristol, Bristol, BS8 1TS, United Kingdom*
[3]*Cavendish Laboratory, Cambridge University, Cambridge, CB3 0HE, United Kingdom*

We show that the interaction potential between sterically stabilized, nearly hard-sphere [poly(methyl methacrylate)–poly(lauryl methacrylate) (PMMA-PLMA)] colloids at a water-oil interface has a negligible unscreened-dipole contribution, suggesting that models previously developed for charged particles at liquid interfaces are not necessarily applicable to sterically stabilized particles. Interparticle potentials, $U(r)$, are extracted from radial distribution functions [$g(r)$, measured by fluorescence microscopy] via Ornstein-Zernike inversion and via a reverse Monte Carlo scheme. The results are then validated by particle tracking in a blinking optical trap. Using a Bayesian model comparison, we find that our PMMA-PLMA data is better described by a screened monopole only rather than a functional form having a screened monopole plus an unscreened dipole term. We postulate that the long range repulsion we observe arises mainly through interactions between neutral holes on a charged interface, i.e., the charge of the liquid interface cannot, in general, be ignored. In agreement with this interpretation, we find that the interaction can be tuned by varying salt concentration in the aqueous phase. Inspired by recent theoretical work on point charges at dielectric interfaces, which we explain is relevant here, we show that a screened $\frac{1}{r^2}$ term can also be used to fit our data. Finally, we present measurements for poly(methyl methacrylate)–poly(12-hydroxystearic acid) (PMMA-PHSA) particles at a water-oil interface. These suggest that, for PMMA-PHSA particles, there is an additional contribution to the interaction potential. This is in line with our optical-tweezer measurements for PMMA-PHSA colloids in bulk oil, which indicate that they are slightly charged.

## I. INTRODUCTION

The interaction between particles adsorbed to a liquid interface, and therefore their microstructure, affects the rheological properties of that interface [1]. These properties play a role in the formation and stability of systems with large interfacial area, such as particle-stabilized emulsions and foams [1–3], which have well-known and widely used applications in the personal care, mineral, and food sectors [4–7]. Understanding the interparticle interaction is therefore important to understand the properties of Pickering systems.

Previous work has considered the microstructure and interactions of charge stabilized particles at liquid-air or liquid-liquid interfaces. Pieranski [8] showed that, for polystyrene particles at a water-air interface, the interaction can be described by a long range dipole-dipole interaction. Further work showed that a combination of a screened Coulomb potential and a long range dipole-dipole interaction gave a more complete description [9,10],

$$U(r) = \frac{A}{r}e^{-\kappa r} + \frac{B}{r^3}, \qquad (1)$$

where $A$ and $B$ are prefactors related respectively to the charge and the effective dipole moment of the particles, $\kappa$ is the screening length in water, and $r$ is the separation between two particles. More recently, studies on polystyrene particles at oil-water interfaces [11,12] concluded that the colloidal repulsion observed there might be due to either residual charges on the oil side [11] or charges on the water side [12] of the particle.

In contrast, there has been less work investigating the nature of the interaction between *sterically* stabilized interfacial particles, which can behave as nearly hard spheres [13]. Like charge stabilized particles, sterically stabilized colloids can be used to stabilize large interfaces. A common particle choice is poly(methyl methacrylate) (PMMA) with polymer hairs grafted to the surface to prevent aggregation due to van der Waals forces [13,14]. PMMA stabilized with poly(12-hydroxystearic acid) (PHSA) is often used in dodecane as a model hard sphere system [15,16], although it has recently been noted that when these particles attach to a dodecane-water interface the particles appear to show a long range repulsion [17]—the origin of this repulsion is unclear as these particles have been shown to behave as hard spheres in dodecane [15,16] and are not stable in water. Additionally, it was found that PMMA-PHSA particles display a dipole-dipole repulsion on interfaces between water and a cyclohexyl bromide (CHB)-alkane mixture, which is prone to light-induced dissociation [18,19]. It was suggested that this arises because PMMA-PHSA particles suspended in the CHB component acquire an effective charge—this is consistent with them forming colloidal crystals with large lattice spacings in this solvent [20].

---

*j.h.j.thijssen@ed.ac.uk

In the present work, we first investigate the long range interaction of PMMA particles sterically stabilized with poly(lauryl methacrylate) (PLMA), as these behave as hard spheres in oil and are unlikely to acquire charge in water. We find that models previously developed for particles at liquid interfaces are not applicable; specifically we show that our data is better described by a screened monopole only rather than Eq. (1), and propose a new model for the long range interaction observed. We use two methods to find the pair potential, $U(r)$, for interfacial PMMA particles. First, we measure radial distribution functions $g(r)$ from fluorescence micrographs, which we convert to pair potentials using an Ornstein-Zernicke inversion; we also fit $g(r)$ by a reverse Monte Carlo method. Second, we employ a blinking optical trap to measure $U(r)$ between interfacial particle pairs. Our data suggest a negligible dipole-dipole contribution and are better fitted with a screened Coulomb potential only. We also find that this interaction can be tuned by introducing salt in the water phase, which lessens the repulsive interaction between the particles. The negligible dipole component we measure suggests that previous models developed for charge stabilized particles at liquid interfaces [11,12] are inapplicable to our system. We instead attribute the long range interaction to the repulsion between neutral holes on a homogeneously charged interface. We also present measurements for PMMA-PHSA particles at a water-oil interface, as these particles are widely used in the literature as near hard spheres. We find that PMMA-PHSA colloids in bulk dodecane are slightly charged and that this leads to an additional term in the interaction potential when attached to an interface, especially at large separations. Finally, we compare our experimental data for PMMA-PLMA to recent theoretical results for the interaction between point charges at a dielectric interface, which we argue are relevant here.

## II. EXPERIMENTAL METHODS

For the experiments reported here, we used two types of colloidal particles: PMMA stabilized with PLMA [poly(lauryl methacrylate)] with diameter 2.4 $\mu$m and polydispersity of 2.5% (determined by Static Light Scattering, SLS) (synthesized following [21]) and PMMA stabilized with PHSA with diameter 2.2 $\mu$m and polydispersity 2.4% (SLS) (synthesized following [15]). These are referred to as PMMA-PLMA and PMMA-PHSA particles, respectively. For the measurement of $g(r)$ at low surface fraction [Fig. 2(c)], PMMA-PLMA with diameter 3.0 $\mu$m and polydispersity 5% was used. The PLMA has a radius of gyration of 2.5 nm in good solvent (n-dodecane; Acros organics, 99%) from dynamic light scattering (DLS), while the PHSA has a radius of gyration of 2.6 nm from DLS and an end-to-end distance of 19 nm when grafted to the colloid surface [22]. PMMA-PLMA has a contact angle of 123° at the water-oil interface (determined by a light extinction technique, LE [23]) and PMMA-PHSA has a contact angle of 121° at the water-oil interface (LE).

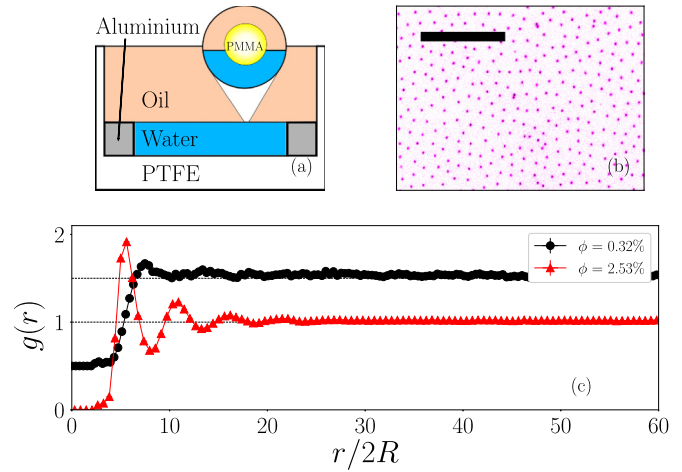

FIG. 1. (a) Schematic of a colloidal particle at a liquid-liquid interface. (b) Experimental micrograph (PMMA-PLMA) showing the structure of these colloidal particles when adsorbed to an interface (zoomed in and inverted from original image for clarity). Gravity points into the page. Scale bar is 100 $\mu$m. (c) The radial distribution function, $g(r)$, of interfacial PMMA-PLMA particles (radius $R = 1.5$ $\mu$m) extracted from a series of these micrographs at two different particle surface coverages, $\phi$, 0.32% and 2.53%. The lower surface fraction is shifted vertically by 0.5 for clarity. Errors in $g(r)$ are of the same order as the symbol size.

All particles were kept as dispersions in n-dodecane which had been filtered three times through an alumina column to remove polar impurities. Distilled and deionized water (Milli-Q, resistivity 18 M$\Omega$ cm) was used as the subphase in all interfacial experiments. We used sodium chloride solutions to perform measurements with a salt solution subphase at 0.01 M, 0.1 M, and 1.0 M.

All interfaces were prepared using the same method. A small polytetrafluoroethylene well was filled with water to a sharp aluminium ledge in order to pin the interface. Above the water layer, 3 ml of low volume fraction dispersion ($\lesssim$0.005%) of PMMA in dodecane was gently spread over the water layer and the flat part of the pinning ledge; see Fig. 1(a). This setup was left for 1–2 h to allow the particles to settle at the interface.

A fluorescence microscope (Nikon Eclipse E800, 10× 0.3 NA objective) was used to take at least 600 snapshots of the interface at an interval of 1 s—an example snapshot is shown in Fig. 1(b). The radial distribution function, $g(r)$, was found from these images using PYTHON code written in-house. Enough snapshots were taken such that the noise in $g(r)$ (quantified by the standard deviation) was $\leqslant$0.03 at large separations $r$ [where $g(r)$ itself is $\sim$1].

We also measure the interparticle potential using a direct method, both in bulk dodecane and at the oil-water interface. A dilute layer (surface coverage $\ll$1%) of particles was adsorbed onto the oil-water interface, and two particles were trapped using a blinking optical trap (BOT) with a power of 0.46 W and a wavelength of 1064 nm (diode pumped

Nd:YAG laser, IPG photonics). The particles were brought to a separation where the potential is expected to be small. The optical trap then blinked on and off at a frequency of 20 Hz. During the time that the lasers were off, the particles' motions were tracked and the diffusion coefficient and speeds were measured from mean squared displacement (MSD) vs time and displacement vs time plots. The force was then calculated using the Stokes-Einstein relation

$$F = \frac{k_B T v}{D}, \quad (2)$$

where $v$ is the speed, $D$ is the diffusion coefficient, and $k_B T$ is the thermal energy. This was repeated at closer and closer separations. Interparticle potentials were then calculated via a numerical integration using the cumulative trapezoidal method.

Zeta potential measurements were performed using a Malvern Nano-Z Zetasizer on a mixture of dodecane and water in a ratio of 1:9. The mixture was emulsified by a vortex mixer for 1 min with no stabilizer present before measuring the zeta potential using a dip cell.

The research data presented in this publication are available on the Edinburgh DataShare repository [24].

## III. RESULTS AND DISCUSSION

From the radial distribution function shown in Fig. 1(c) we can see long range order in this system, with measurable correlations persisting up to ∼20 particle diameters. The source of this repulsion is unclear, with PMMA-PLMA acting as a hard sphere in dodecane (see below) and unlikely to acquire charge in water (given it has no dissociable groups).

We start our quantitative analysis by considering the measurements for interfacial PMMA-PLMA particles. $g(r)$ for PMMA-PLMA particles at the water-oil interface at low surface fraction (0.32%), shown in Fig. 1(c), were converted to pair potentials, $U(r)$, using an Ornstein-Zernicke (OZ) inversion with the Percus-Yevick (PY) approximation [25,26]. We fit the pair potential to Eq. (1) as well as to a single screened Coulomb potential [i.e., only the first term in Eq. (1)]; the results are summarized in Table I. We see that the dipole contribution is negligible, which we can quantify by observing the dimensionless quantity $\frac{B\kappa^2}{A} \ll 1$. The almost identical values of reduced $\chi^2$ indicate that both models fit the data comparably, with the extra fit parameter due to the second term in Eq. (1), leading to a slightly less favorable fit.

We also wish to find interparticle potentials at higher density, where the OZ inversion becomes less reliable [26]. With this in mind, the $g(r)$ were also inverted to $U(r)$ via a reverse Monte Carlo scheme in order to obtain fit parameters at a higher surface fraction (2.49%), as OZ inversion only works reliably at low surface fraction [26]. Analyzing experimental data obtained at higher particle surface fractions is beneficial as it probes smaller interparticle separations. A parametrized pair potential was used to run a Monte Carlo simulation and $g(r)$ was extracted from the results. The parameters were then varied to find an optimum fit, corresponding to a minimum in a normalized $\chi_g^2$ parameter. Given the results of our OZ inversion (Table I), we use a screened monopolar potential as our parametrization. The parametrized potential and form of $\chi_g^2$ are

$$U(r) = \frac{A}{r} e^{-\kappa r}, \quad \chi_g^2 = \frac{1}{N} \sum_{i=1}^{N} \frac{\left[g_{\text{expt}}^{(i)}(r) - g_{\text{sim}}^{(i)}(r)\right]^2}{\Delta_i^2}, \quad (3)$$

where $\Delta_i$ is the measured error on point $i$, and there are $N$ such points. Using this reverse Monte Carlo scheme, the pair potential for PMMA-PLMA has been obtained. Figure 2 shows the results of this method. A set of parameters providing good fits are $A \simeq 1964\ k_B T\ \mu$m and $\kappa \simeq 0.38\ \mu$m$^{-1}$ [Figs. 2(b) and 2(c)].

TABLE I. Summary of the results of fitting $U(r)$ from an OZ inversion, using the PY closure relation, to experimental data. D refers to a screened monopole plus a dipole [Eq. (1)], while M refers to just a screened monopole, i.e., just the first term of Eq. (1). The reduced $\chi^2$ statistic assumes the same error on each data point and includes division by the number of degrees of freedom.

| Model | $A/k_B T\ \mu$m | $\kappa/\mu$m$^{-1}$ | $B/k_B T\ \mu$m$^3$ | Reduced $\chi^2$ |
|---|---|---|---|---|
| D | 1036 | 0.27 | $1.36 \times 10^{-16}$ | 0.0054 |
| M | 1036 | 0.27 | | 0.0053 |

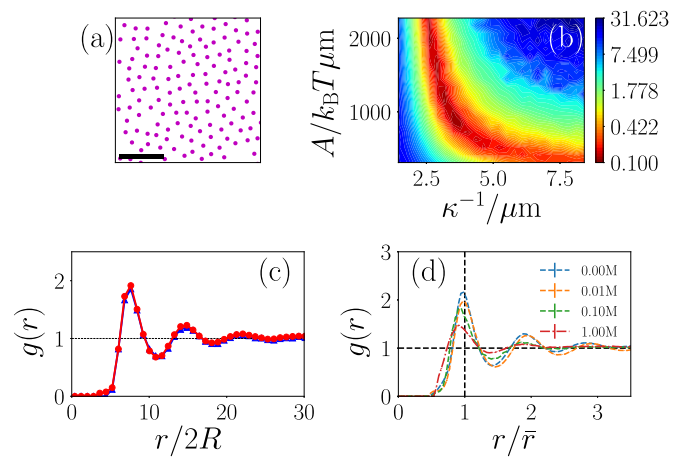

FIG. 2. (a) Simulated snapshot of particles at an interface; scale bar is 100 $\mu$m. (b) Contour plot of $\chi^2$ as a function of $\kappa^{-1}$ and $A$ for PMMA-PLMA. Optimal fits are minima in this plot. (c) Comparison of experimental (red line, ◯) and simulated (blue line, △) $g(r)$ for PMMA-PLMA particles at $\phi = 2.53\%$; the experimental and simulated line overlap visually. (d) Plot of $g(r)$ for PMMA-PLMA at an oil-water interface at various salt concentrations. $r$ is scaled by $\bar{r}$, which is the average interparticle separation based on surface coverage. Surface fractions are 0.00 M—3.40%, 0.01 M—3.23%, 0.10 M—4.04%, and 1.00 M—2.92%.

A few remarks are in order at this point. First, there are multiple values of $(A, \kappa)$ which provide similar values of $\chi_g^2$ [Fig. 2(b)]—the order of magnitude is the same though. For instance, the minimum in $\chi_g^2$ occurs at (1964 $k_B T\ \mu$m, 0.38 $\mu$m$^{-1}$) with $\chi_g^2 = 0.1102$, whereas $\chi_g^2 = 0.1110$ at (931 $k_B T\ \mu$m, 0.29 $\mu$m$^{-1}$), close to the values obtained from OZ inversion. This is expected as phase behavior should largely depend on the second virial coefficient (rather than on $A$ and $\kappa$ separately). Second, the relatively large value of the Debye screening length, $\kappa^{-1} \sim 3\ \mu$m, implies that the interaction propagates, at least in part, through the oil phase as water has a maximum Debye length of $\sim 1$ $\mu$m at very low ionic strengths [27,28].

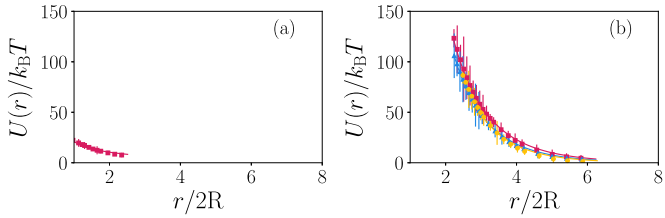

FIG. 3. Energy profiles for PMMA-PLMA when in bulk dodecane (a) and adsorbed to a dodecane-water interface (b) measured with a blinking optical trap. $r$ is core-to-core separation and $R$ is the particle radius; different symbols/colors correspond to different particle pairs.

The interparticle potentials for PMMA-PLMA measured using the BOT are shown in Fig. 3. The energy curves were fit to a screened Coulomb interaction [29] given in Eq. (3). The prefactor is

$$A = \frac{Q_{\text{eff}}^2}{4\pi\epsilon\epsilon_0}, \qquad (4)$$

where $\epsilon$ is taken to be the average permittivity of the two phases and $\kappa$ is the inverse screening length. We see from Fig. 3(a) that, as expected, our PMMA-PLMA particles only have a relatively small force in dodecane, approaching hard-sphere-like behavior. Moreover, the chemical structure of the PLMA stabilizer does not appear to feature any dissociable groups, so there is no obvious mechanism for these particles to acquire charge in water. This suggests that the long-range interaction between interfacial PMMA-PLMA particles observed here is indeed due to the liquid interface.

For the interfacial case [Fig. 3(b)], the optimal fit parameters were found to be $A = 3400 \pm 900$ $k_B T$ $\mu$m and $\kappa = 0.310 \pm 0.008$ $\mu$m$^{-1}$. From this, we can calculate the effective charge, $Q_{\text{eff}}$, and find the surface charge density of the oil-water interface, $\sigma = \frac{Q_{\text{eff}}}{\pi R^2 \sin^2(\theta)}$. Doing this we find $\sigma = 7.8 \pm 0.9$ nC cm$^{-2}$. Using the prefactor value obtained from inverting $g(r)$ we obtain $\sigma = 5.9 \pm 0.6$ nC cm$^{-2}$, which is in fair agreement. Differences between BOT and $g(r)$ inversion results can be attributed to the heterogeneity of the interparticle interaction between different particle pairs, where $g(r)$ inversion involves the entire ensemble, whereas the BOT experiment relies on specific pairs of particles. Park *et al.* note that particle pair interactions at the lower end of the distribution have a disproportionate effect on the structure of the ensemble, leading to $g(r)$ measurement techniques consistently finding apparently weaker interaction strengths [30].

To determine whether the interfacial BOT energy curves are better described by a functional form having a screened Coulomb plus dipole term, Eq. (1), or a screened monopole term only, Eq. (3), we performed a Bayesian model comparison [31]. This analysis shows that the posterior probability ratio for each curve is $\sim 40$, in favor of the model with a screened monopole term only (with an effective screening length; see Appendix A). The same Bayesian method was used to compare a screened monopole term to other possible models, for example, screened dipole or screened monopole plus screened dipole, which gave similar values of the posterior probability ratio, i.e., of order 40. These analyses show that a single screened Coulomb potential with an effective screening length provides a decent fit to our data, and the best fit of the models tested here, in line with our numerical solution to the interaction between point charges at a dielectric interface (see Appendix A).

Based on the hard sphere behavior of PMMA-PLMA particles and the consideration that PLMA has no mechanism to acquire charge in water, previous theories for charged particles at liquid interfaces [11,12] are not directly applicable. For these reasons, we propose an alternative model based on the idea of neutral holes in a charged plane. It is known that water-alkane interfaces can become charged [32–35] and, further, our measurements of the zeta potential of dodecane droplets in water give $-65 \pm 13$ mV, which is in line with measurements made by Marinova *et al.* [32] and Creux *et al.* [34]. Invoking superposition at the level of the Poisson equation, we can consider that an array of neutral holes on a charged sheet will behave as an array of charged holes on a neutral sheet as far as in-plane interactions are concerned (we neglect the homogeneous electric field perpendicular to the interface as it does not contribute to the pair interaction). The holes will have an effective charge given by $Q_{\text{eff}} = a\sigma$, where $a$ is the cross-sectional area of the particle at the interface and $\sigma$ is the surface charge density of the bare liquid interface.

The interaction between the effectively charged holes can be found by solving an interfacial Poisson-Boltzmann equation. The solution is obtained by using the methods in [9,10], and may be approximated at relatively short distances by a screened Coulomb potential. Moreover, a single screened Coulomb potential with a modified effective screening length provides a decent fit to the data over a large range of separations $r$ (see Appendix A).

If the repulsion we observe is caused by neutral holes existing in a charged plane, it is fundamentally an electrostatic one. We therefore tested the effect on the interaction of adding salt to the water phase. Comparing $g(r)$ measurements at comparable surface coverage shows a decrease in order upon increased aqueous salt concentration, quantified by a lowering of the first peak height as salt concentration is increased; see Fig. 2(d) (see Appendix B). This decrease in order can be explained in one of two ways: a decrease in the effective charge or a decrease in the screening length.

A change in effective charge could be achieved by a changing contact angle as the area blocked by the particle is given by $A = \pi R^2 \sin^2(\theta)$. A light extinction technique to measure contact angle [23] was used to check this and it was found that, for PMMA-PLMA, there is no apparent dependence of contact angle on salt concentration. As there is no evidence for the salt dissolving in the oil phase, and so we expect no change in screening in the oil phase, the salt must directly lower the interfacial charge itself. This conclusion is in line with measurements of a lower absolute value of zeta potential of the bare water-dodecane interface upon salt addition by Marinova *et al.* [32] and Creux *et al.* [34]. Note that, for the case of charge-stabilized particles, a relatively weak dependence of interfacial particle interactions on aqueous salt concentration can be explained by nonlinear charge renormalization at the particle-water surface [36]. However, this model also predicts a $\frac{1}{r^3}$ dependence of the interaction potential, which does not align with our data.

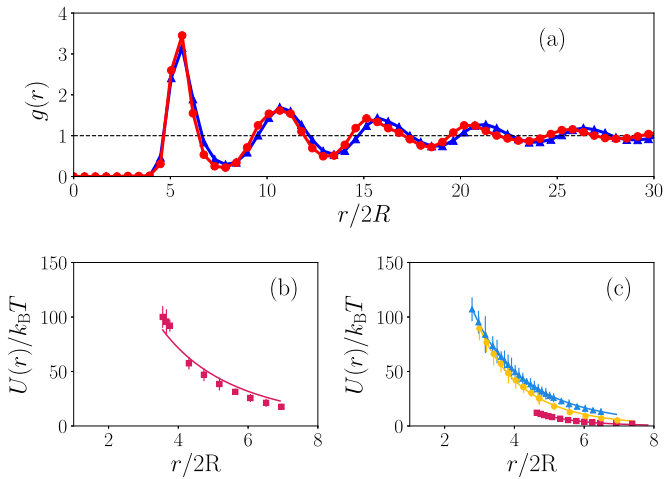

FIG. 4. (a) Comparison of experimental (red line, ◯) and simulated (blue line, △) $g(r)$ for PMMA-PHSA particles at $\phi = 2.19\%$; the experimental and simulated line overlap visually. (b), (c) Energy profiles for PMMA-PHSA when in bulk dodecane (b) and adsorbed to a dodecane-water interface (c) measured using a blinking optical trap. $r$ is core-to-core separation and $R$ is the particle radius; different symbols/colors correspond to different particle pairs.

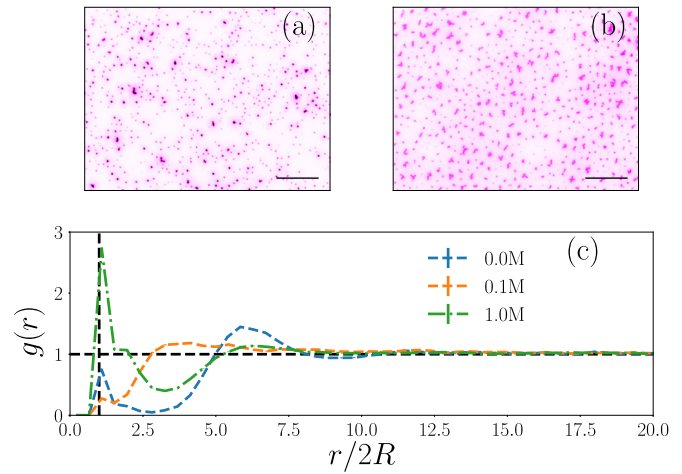

FIG. 5. (a), (b) Snapshots of a collection of PMMA-PHSA colloids at a dodecane-salt solution interface, at 0.1 M (a) and 1.0 M (b) NaCl, showing aggregation upon addition of salt. Surface coverages were 2.45% (for the 0.1 M case) and 3.6% (for the 1 M case). (These surface fractions lead to similar number fractions of aggregates plus individual particles.) The scale bars are both 100 $\mu$m. (c) Measured $g(r)$ for PMMA-PHSA particles at a water-oil interface where NaCl has been added to the water phase at different concentrations.

We have also performed both the $g(r)$ and the BOT experiments on water-dodecane interfaces laden with PMMA-PHSA particles, as these have been widely used in the literature as near hard spheres. In Fig. 4(a), we observe that the parametrized potential [Eq. (3)] provides a decent fit, with fit parameters $A \simeq 4136 \, k_BT \, \mu$m and $\kappa \simeq 0.35 \, \mu$m$^{-1}$. It is worth noting that this value for $A$ is within error of that obtained for PMMA-PLMA particles, whereas the $g(r)$ of PMMA-PLMA and PMMA-PHSA are markedly different. We see that a moderate change to the interaction potential results in a considerable change (at least qualitatively) to the $g(r)$. However, it is clear that this fit is not as good as that seen for PMMA-PLMA in Fig. 2(c). This suggests that there might be additional contributions to the interaction between interfacial PMMA-PHSA particles, i.e., beyond the interaction between neutral holes in a charged plane.

The BOT data in Fig. 4(b) reveals that the PMMA-PHSA particles do not behave as hard spheres in bulk dodecane (also see [37]). The long range force that we measure can be fit with a screened monopole [Eq. (3)] leading to optimal fit parameters of $A = 1400 \pm 70 \, k_BT \, \mu$m and a decay length, $\kappa^{-1}$, of $10.9 \pm 1.1 \, \mu$m. Using the bulk equivalent of Eq. (4) we find that the particles have a surface charge density of $2.3 \times 10^{-4} \, \mu$C cm$^{-2}$. It is worth noting that this surface charge density is four orders of magnitude lower than the particles used by either Aveyard et al. [11] or Masschaele et al. [12]. However, when fitting the BOT data at the interface [Fig. 4(c)], the Bayesian model comparison indicates that these data are ∼10 times more likely to be described by a screened monopole than a combination of a screened monopole and dipole.

These considerations imply that for PMMA-PHSA the potential is more complicated as the missing area repulsion is complemented by an additional repulsion between surface charges, which are in general off the liquid interface. This explanation is in line with the $g(r)$ data in Fig. 4(a), where the simulated $g(r)$ and experimental $g(r)$ begin to diverge from each other at larger $r$.

We also tested the effect on the interaction of PMMA-PHSA of adding salt to the water phase. Figure 5 shows the results for PMMA-PHSA of salt addition resulting in 0.1 M and 1 M solutions as well as with no salt added in Fig. 5(c). We observe aggregation [38] into colloidal clusters of self-limiting size (microphase separation). This can be explained as follows. Initially, the salt reduces the electrostatic repulsion, allowing capillary and van der Waals interactions to facilitate aggregation (see below for a more quantitative discussion of these). As the aggregates grow, the area of interface blocked by that aggregate increases and therefore so does the effective charge of that neutral hole. We therefore observe aggregates eventually behaving as larger, interfacially adsorbed particles which have their own long range repulsion and order. Such aggregates of self limiting size have been observed previously for charge stabilized colloidal systems [39].

So far, we have not estimated possible sources of attraction between the particles. As in the bulk, we expect van der Waals forces should be counteracted by the steric stabilizer, especially given that the majority of the particle sits in the oil phase ($\theta \simeq 120°$). Capillary forces, however, may be present. The Bond number gives the ratio of gravitational to surface tension effects, Bo $= R^2 \Delta \rho g / [\gamma (1 − \cos \theta)]$, where $\Delta \rho$ is the density difference between the particle and the lower phase, $g$ is acceleration due to gravity, $\gamma$ is the interfacial tension, and $\theta$ is the contact angle [40]. For our particles Bo $\sim 10^{-8} \ll 1$, indicating that gravitational effects are negligible and therefore there should be no flotation capillary forces [41]. Surface roughness, however, due to the polydispersity in stabilizer length, could induce capillary attractions, which may cause attraction when the electrostatic repulsion is suppressed by the addition of salt to the water phase [42].

The analysis of our experimental data so far has been based on Eq. (1), suggesting that the interaction between sterically stabilized particles at a water-oil interface can be described by the interaction of neutral holes in a charged liquid interface only, particularly in the case of PMMA-PLMA. We have also argued that, as far as in-plane interactions are concerned, this is equivalent (at the level of the Poisson equation) to the interaction of disks of charge $Q_{\text{eff}}$ on a neutral interface. Replacing the charged disks with point charges $Q_{\text{eff}}$ at the centers of the disks, this should be equivalent to the interaction between point charges at a dielectric interface, which has recently been solved analytically [43]. The theoretical result can be described as a single screened Coulomb potential (with $\kappa_{\text{water}} = 10\kappa_{\text{oil}}$) crossing over to $\frac{e^{-\kappa_{\text{oil}}r}}{r^2}$. As the crossover distance for our experiments is estimated to be 10 $\mu$m [43], we reanalyze our PMMA-PLMA data by fitting the following functional form

$$U(r) = A \frac{e^{-\kappa_{\text{oil}}r}}{r^2} \quad (5)$$

to our $g(r)$ data for PMMA-PLMA, where most of the data is in the region $r > 10$ $\mu$m.

Numerically, a screened monopole provides a (marginally) better fit than Eq. (5), but the latter provides fitting parameters that are physically more consistent. For example, screened-monopole fits result in values for $A$ and $\kappa$ that change nonmonotonically with increasing salt concentration in the aqueous phase. On the contrary, fits using Eq. (5) result in $6.0 < \kappa_{\text{oil}}^{-1} < 6.5$ $\mu$m and a monotonically decreasing value for $A$ upon increasing salt concentration from 0 to 1.0 M. Notably, $\kappa_{\text{oil}}^{-1} = 6.5$ $\mu$m is closer to the decay length of 10.9 $\mu$m that we obtained in bulk dodecane than the $\sim$3 $\mu$m from the screened-monopole fit in the no-salt case, and the decrease in $A$ corresponds with a similar trend observed in our measurements upon salt addition [Fig. 2(d)]. At high salt concentration, i.e., 0.1 M and 1.0 M, fits using Eq. (5) are less good (see Appendix B), which is in line with the emergence of a screened-dipole regime in the theory [43].

In addition, fits to the BOT data for PMMA-PLMA using Eq. (5) result in similar fit parameters to the fits to our $g(r)$ data for PMMA-PLMA. Our Bayesian analysis indicates that, when fitting every data point, Eq. (3) provides a better fit, whereas if we fit for $r > 7$ $\mu$m only, Eq. (5) provides a better fit. Therefore, we can say that our data for PMMA-PLMA can be better described by either a screened monopole or by the form given in Eq. (5) rather than a dipolar fit.

## IV. CONCLUSION

In conclusion, we have experimentally shown that sterically stabilized, nearly hard-sphere PMMA-PLMA particles exhibit a long range repulsion when attached to an oil-water interface. We have also demonstrated that this interaction can be altered by changing the salt concentration in the aqueous phase. Quantitatively, the long-range repulsion observed has a negligible unscreened-dipole contribution; instead our data is better described by a screened Coulomb potential with an effective screening length. We attribute this long-range interaction to the particles acting as neutral holes in the charged plane of the water-oil interface. Hence we have also fitted our data to recent theoretical results for the interaction between point charges at a dielectric interface. This fit is marginally worse than the screened-monopole case, but it provides fitting parameters that are physically more consistent, especially when considering the addition of salt to the aqueous phase.

We have also presented measurements for PMMA-PHSA particles at a water-oil interface. At relatively small interparticle separations $r$, the data is consistent with the interaction between neutral holes in a charged plane. At larger $r$, the data suggests that an additional contribution to the interaction potential is required, which is in line with our optical-tweezer measurements that indicate that our PMMA-PHSA particles are slightly charged in bulk dodecane.

The generic point of our results is that, while existing models for particles at liquid interfaces consider the charge at the particle-water and/or particle-oil surfaces [11,12], the charge of the liquid interface cannot, in general, be ignored. Notably, this statement applies to any Pickering system where the fluid-fluid interface has a charge, not an unlikely scenario given [32–35]. Finally, to provide a further test for our neutral hole explanation, future experiments could focus on varying the charge density of the liquid interface, for example, by changing the pH and the salt concentration in the aqueous phase in a controlled manner, so as to keep the ionic strengths (and hence the Debye lengths) in the two phases constant.

## ACKNOWLEDGMENTS

I.M. acknowledges studentship funding from the EPSRC Centre for Doctoral Training in Condensed Matter Physics (CM-DTC, EP/L015110/1). J.H.J.T. acknowledges The University of Edinburgh for support through a Chancellor's Fellowship. We acknowledge A. Law and D. M. A. Buzza for useful discussions, L. Kobayashi Frisk for preliminary measurements, T. Wallner for preliminary measurements and preliminary data analysis, and M. Turley and J. Arlt for help in setting up contact angle measurements.

## APPENDIX A: NUMERICAL SOLUTION OF AN INTERFACIAL POISSON PROBLEM

If the interaction between nearly hard colloidal spheres at a water-oil interface does indeed stem from the liquid-liquid interface being charged, and that is equivalent to charged disks in a neutral interface, then we can model the interaction as that between two point charges (at the center of the disks) at an interface. The potential between two charged particles (each of charge $Q$) at an interface between two media (in our case water and dodecane), with dielectric constants $\epsilon_1$ and $\epsilon_2$, respectively, can be worked out by following [9,10]. The dielectric constants are the product of the vacuum permittivity $\epsilon_0$ and the relative permittivity of the medium. The potential

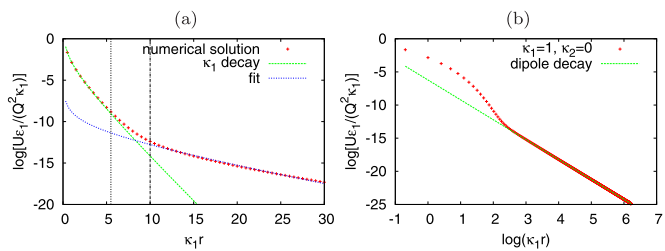

FIG. 6. Potential $U(r)$ between two charges at the interface between two dielectric media. (a) Plot of $U(r)$ for $\kappa_1 = 10\kappa_2$. This curve gives a prediction of what we expect for our water-dodecane system. At relatively small distances $r$, the numerical data are well approximated by a $\frac{Q^2}{4\pi\epsilon_1}\frac{\exp(-\kappa_1 r)}{r}$ functional form ($\kappa_1$ decay). The dashed line indicates the minimum distance probed using the blinking optical trap technique and the dot-dashed line indicates the minimum distance probed using the $g(r)$ inversion technique, both assuming $\kappa_1 = 1\ \mu m^{-1}$. The "fit" label in the legend refers to a single screened Coulomb fit with an effective decay constant $\kappa$, $A\frac{\exp(-\kappa r)}{r}$. For our numerics, we get $\kappa \simeq 0.18\kappa_1$ and $A \simeq 1.7 \times 10^{-4}Q^2/\epsilon_1 = Q^2_{\text{eff}}/(4\pi\bar{\epsilon})$, with $Q \simeq 30Q_{\text{eff}}$. (b) Plot of $U(r)$ for $\kappa_1 = 1$ and $\kappa_2 = 0$, showing an asymptotic dipole behavior [$U(r) \propto 1/r^3$]. This curve gives a prediction of what would be expected for an air-water interface, as originally considered in [9,10].

is (in SI units)

$$U(r) = \frac{Q^2}{4\pi\bar{\epsilon}r}y(\kappa_1 r, \kappa_2 r; \epsilon_1, \epsilon_2), \quad (A1)$$

where $\bar{\epsilon} = \frac{\epsilon_1+\epsilon_2}{2}$, and with

$$y(\kappa_1 r, \kappa_2 r; \epsilon_1, \epsilon_2) = \int_0^{+\infty} dx \frac{xJ_o(x)}{\xi_1\sqrt{x^2+k_1^2} + \xi_2\sqrt{x^2+k_2^2}}, \quad (A2)$$

where $k_{1,2} = \kappa_{1,2}r$, $\xi_{1,2} = \epsilon_{1,2}/(\epsilon_1+\epsilon_2)$, and $J_0$ is the zeroth order Bessel function of the first kind. Note that $\kappa_1^{-1}$ and $\kappa_2^{-1}$ are the Debye length in the first and second phase (here water and dodecane, respectively).

Figure 6 shows the potential $U(r)$ in Eq. (A1) calculated for (a) $\kappa_1 = 10\kappa_2$ and $\epsilon_1 = 40\epsilon_2$, as relevant for our measurements, and (b) for $\kappa_2 = 0$ and $\epsilon_1 = 80\epsilon_2$, as relevant for an air-water interface. For the case in (a), we note that the potential may be approximated at relatively short distances by a screened Coulomb potential with a screening length of $\kappa_1^{-1}$ ($\kappa_1$ decay). At sufficiently large distance, there is a crossover to another exponentially decaying behavior with decay constant equal to $\kappa_2^{-1}$. The exact asymptotic behavior at large $r$ is worked out in [43]. In addition, a single screened Coulomb potential with an effective screening length (in between $\kappa_1^{-1}$ and $\kappa_2^{-1}$) provides a decent fit over a large range [Fig. 6(a)]. These observations explain why a single screened Coulomb potential is sufficient to fit our experimental data.

## APPENDIX B: EFFECTIVE POTENTIAL UNDER DIFFERENT SALT CONDITIONS

In the main text we fitted an effective potential to our $g(r)$ data without salt, assuming a single screened Coulomb potential with adjustable prefactor ($A$) and inverse screening length $\kappa^{-1}$; see Eq. (3) in the main text.

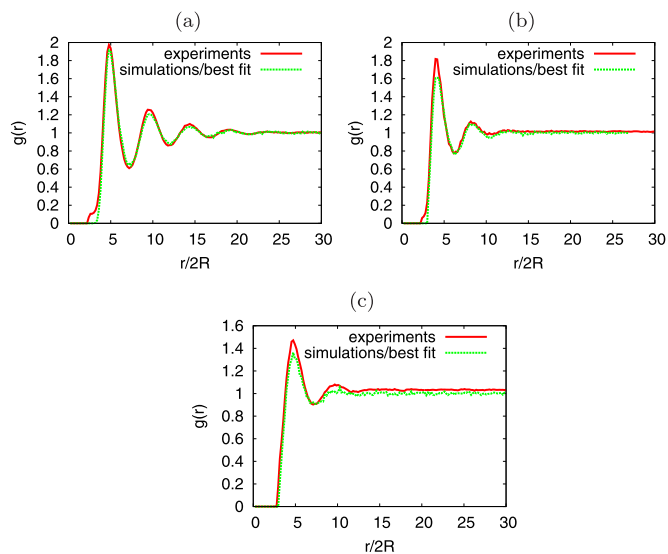

FIG. 7. Fitting potentials for the salt (NaCl) data for PMMA-PLMA with a screened monopole. (a) 0.01 M data: the best fit is obtained for $A \simeq 1861\ k_BT\ \mu m$ and $\kappa \simeq 0.43\ \mu m^{-1}$ in Eq. (3) in the main text. (b) 0.1 M data: best fit for $A \simeq 1654\ k_BT\ \mu m$ and $\kappa \simeq 0.57\ \mu m^{-1}$. (c) 1 M data: best fit for $A \simeq 465\ k_BT\ \mu m$ and $\kappa \simeq 0.41\ \mu m^{-1}$.

Here we repeat the procedure for our salt data. First, we fit again to a screened monopole interaction. The best fits and corresponding parameters are shown in Fig. 7. In all cases a single screened Coulomb potential provides a decent fit of the data, although the fit is comparatively worse for the larger salt concentrations.

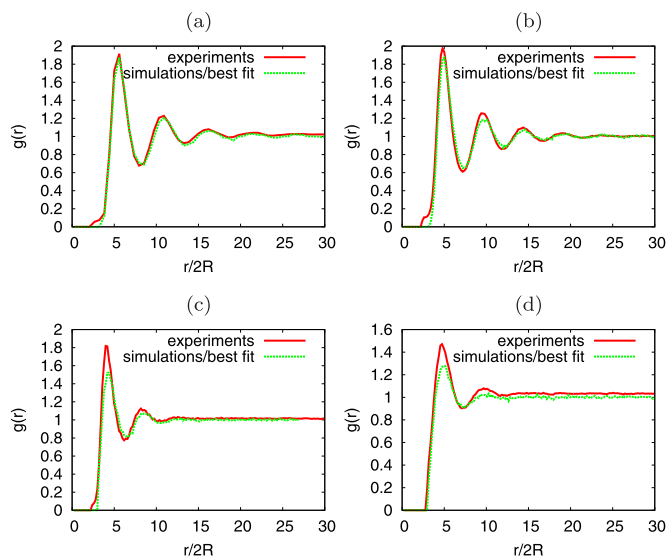

FIG. 8. Fitting potentials for $g(r)$ data without and with salt (NaCl) for PMMA-PLMA with $U = Ae^{-\kappa r}/r^2$. (a) No salt: the best fit is obtained for $A \simeq 3200\ k_BT\ \mu m^2$ and $\kappa \simeq 0.17\ \mu m^{-1}$. (b) 0.01 M data: the best fit is obtained for $A \simeq 2000\ k_BT\ \mu m^2$ and $\kappa \simeq 0.15\ \mu m^{-1}$. (c) 0.1 M data: best fit for $A \simeq 800\ k_BT\ \mu m$ and $\kappa \simeq 0.17\ \mu m^{-1}$. (d) 1 M data: best fit for $A \simeq 600\ k_BT\ \mu m$ and $\kappa \simeq 0.17\ \mu m^{-1}$.

Motivated by the theoretical work in [43], which shows that the asymptotic behavior of Eq. (A1) is $A\frac{e^{-\kappa r}}{r^2}$, we have also used this functional form for fits. The results are shown in Fig. 8. Although the fits are slightly worse than for a single screened Coulomb potential, the value of $\kappa$ obtained from these fits is closer to the inverse Debye length of dodecane, which should be $\sim 0.1\ \mu\mathrm{m}^{-1}$.